\documentclass{elsart}
\usepackage{epsfig}
\begin{document}
\begin{frontmatter}
  \title{Nonequilibrium dynamics of quantum fields}
  \author[label1]{R.L.S. Farias} \author[label1]{N.C. Cassol-Seewald}
  \author[label1]{G. Krein} \author[label2]{R.O. Ramos}
%
  \address[label1]{Instituto de F\'\i sica Te\'orica, Universidade Estadual
    Paulista, \\ Rua Pamplona 145, S\~ao Paulo, SP 01405-900, Brazil}
  \address[label2]{Departamento de F\'{\i}sica Te\'orica, Universidade do
    Estado do Rio de Janeiro, 20550-013 Rio de Janeiro, RJ, Brazil}
\begin{abstract}
  The nonequilibrium effective equation of motion for a scalar background
  field in a thermal bath is studied numerically. This equation
  emerges from a microscopic quantum field theory derivation and it is
  suitable to a Langevin simulation on the lattice.  Results for both
  the symmetric and broken phases are presented.
\end{abstract}

\end{frontmatter}


Relativistic heavy-ion experiments produce highly excited hadronic matter at
high temperatures and densities. One important question in the study of the
properties of the produced excited matter is the understanding of the dynamics
of the quantum fields governing the effective degrees of freedom. In the
present work we will discuss the derivation of stochastic
Ginzburg-Landau-Langevin (GLL) type of equations with additive and
multiplicative noises from quantum corrections to the effective action of a
self-interacting $\lambda \phi^4$ scalar field theory.
We apply the real time Schwinger's closed time path formalism, as explained
for instance in Ref.~\cite{GR}, originally done for the symmetric phase of the
model, and we here extend that application to the broken phase of the model.
Results of numerical lattice simulations of the derived effective GLL equation
in $3$ spatial dimensions are then presented for both the symmetric and broken
phases. Particular attention is devoted to the question of the the
renormalization of the stochastic GLL equation to obtain equilibrium results
that are independent of the lattice spacing.


Consider the scalar field model with action
\begin{eqnarray}
\mathcal{S}[\phi ]= \int d^4 x \left[ \frac{1}{2}\left(
\partial _{\mu }\phi \right)^{2}\mp
\frac{ m^{2}}{2}\phi ^{2}-\frac{\lambda }{4!}\phi^{4} \right]
\end{eqnarray}
where the minus (plus) sign in the quadratic term for the potential is for the
symmetric (broken) phase.  The effective action in terms of {}Feynman
diagrams, up to $O(\lambda^2)$, is given by

\begin{eqnarray}
\!\!\!\!\!\!\!\!\!
\begin{picture}(210,8)
\put(0,-1){$\Gamma[\varphi_c] = S[\phi] \; +$} \thicklines
\put(90,0){\line(-1,-1){10}} \put(100,0){\circle{20}}
\put(90,0){\line(-1,1){10}}
\put(120,-1){+}
\put(150,0){\line(-1,-1){10}} \put(160,0){\circle{20}}
\put(150,0){\line(-1,1){10}} \put(180,0){\circle{20}}
\put(200,-1){+}
\put(225,0){\line(-1,-1){10}} \put(235,0){\circle{20}}
\put(225,0){\line(-1,1){10}} \put(245,0){\line(1,-1){10}}
\put(245,0){\line(1,1){10}}
\put(265,-1){+}
\put(300,2){\circle{20}} \put(280,2){\line(35,0){40}}
\put(325,-1){$+\; O(\lambda^3)$.}
\end{picture}
\label{Gamma}
\end{eqnarray}

{}Following \cite{GR}, the real time Schwinger's closed time path formalism
\cite{chou} is used for obtaining the effective action of the theory. We will
not repeat here the derivation done in \cite{GR}, that can also be carried out
analogously for the broken phase, and only give here the final result for the
effective equation of motion of the order parameter. In the derivation there
appears imaginary terms in the effective action, that can be associated to
functional integrations over Gaussian fluctuation fields $\xi$ by making use
of a Hubbard-Stratonovich transformation. The fields $\xi$ act as fluctuation
sources for the scalar field configuration (order parameter) $\varphi_c
=\langle \phi \rangle$.  In the original calculation of \cite{GR} there
appears two fluctuation (stochastic) fields, $\xi_1$ and $\xi_2$.  $\xi_1$
couples to $\varphi_{c}$, leading to a multiplicative noise term
($\varphi_{c}\xi_1$) in the equation of motion for $\varphi_{c}$, while the
field $\xi_2$ gives origin to an additive noise term.  A Langevin-like
equation for $\varphi_c$ emerges after a series of physically motivated
approximations, related to spatial and temporal nonlocalities, are considered.
{}For instance, considering only slowly-varying contributions in space and
time (this is a valid assumption for systems near equilibrium, when
$\varphi_c$ is not expected to change considerably with time \cite{BGR}), the
final result that we obtain for the effective equation of motion for
$\varphi_c$ is of the form
\begin{equation}
\hspace{-0.5cm}
\left[ \Box + m_{T}^{2}\right] \varphi _{c}\left(
\mathbf{x},t\right) + \frac{\lambda _{T}}{3!}\varphi _{c}^{3}\left(
\mathbf{x},t\right) + \eta\varphi _{c}^{2}\left( \mathbf{x},t\right)
\dot{\varphi}_{c}\left( \mathbf{x},t\right) =\varphi _{c}\left(
\mathbf{x},t\right) \xi _{1}\left( \mathbf{x},t\right),
\label{langevin_equation}
\end{equation}
where $\lambda_T$ and $m_T$ are renormalized finite temperature coupling
constant and the renormalized finite temperature mass.  In the high
temperature approximation, $T/m \gg~1$, and within perturbative values of
$\lambda$, which is the regime we will be exploring in the simulations,
$\lambda_T$ and $m_T$ are well approximated by $\lambda_T \simeq \lambda$ and
\begin{eqnarray}
m_{T}^{2} \simeq \left\{
\begin{array}{ll}
m^{2}+\lambda \frac{T^{2}}{4!} &\hspace{0.5cm}{\rm symmetric \; phase} \\
-m^2+\frac{\lambda}{4!}T^2
+\frac{\lambda}{2}\nu^2 &\hspace{0.5cm}{\rm broken \; phase}
\end{array}
\right.
\end{eqnarray}
where $\nu$ is the vacuum expectation value for the scalar field in the broken
phase \cite{DJack} and $\eta$ is the dissipation coefficient associated with
the multiplicative noise field $\xi_1$ and it is given by
\begin{eqnarray}
\eta \stackrel{T \gg m_T}{\simeq} \frac{96}{\pi T}\ln
\left(\frac{2T}{m_T} \right)
\end{eqnarray}
Thermal corrections can restore the symmetry, the potential can change from
double well to single well for a temperature larger than a critical
temperature $T_c$ given by $T_c^2=m^2/(\lambda/4!)$.  Note that the
additive noise term drops out at $O(\lambda^2)$ \cite{GR}.

The noise treatment involves some care due to appearance of Rayleigh-Jeans
ultraviolet divergences at long times when simulating the equation on a
discrete lattice. These divergences manifest themselves through a
lattice-spacing dependence of the equilibrium solutions. We simulate the GLL
equations discretizing time in steps of $\Delta t$ using a leapfrog algorithm
and treat the spatial coordinate using {}Fast-Fourier transform on a cubic
lattice of side $L$ (as in Ref.~\cite{GF}). In order to illustrate the
dependence of the solutions to the lattice spacing $h = L/N$, where $N$ is the
number of lattice sites in each spatial direction, we show in {}Fig. 1 results
for a simple double-well potential in dimensionless units such that the
temperature is $T = 1$, the dissipation coefficient is $\eta = 1$ and $\Delta
t = 0.001$.

\vspace{0.575cm}
\begin{figure}[ht]
\begin{center}
  \includegraphics[scale=0.2725]{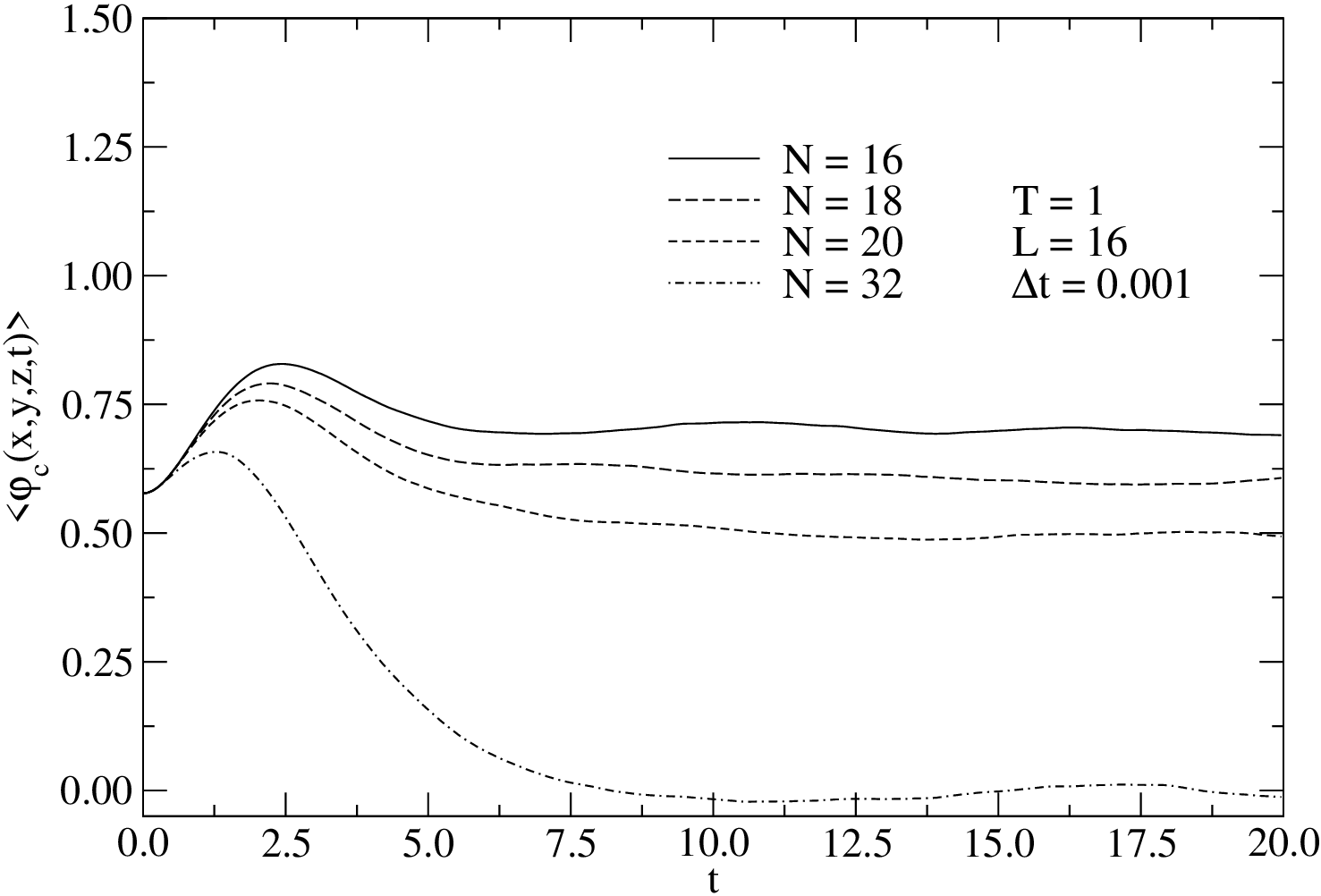}\hspace{0.1 cm}
  \includegraphics[scale=0.2725]{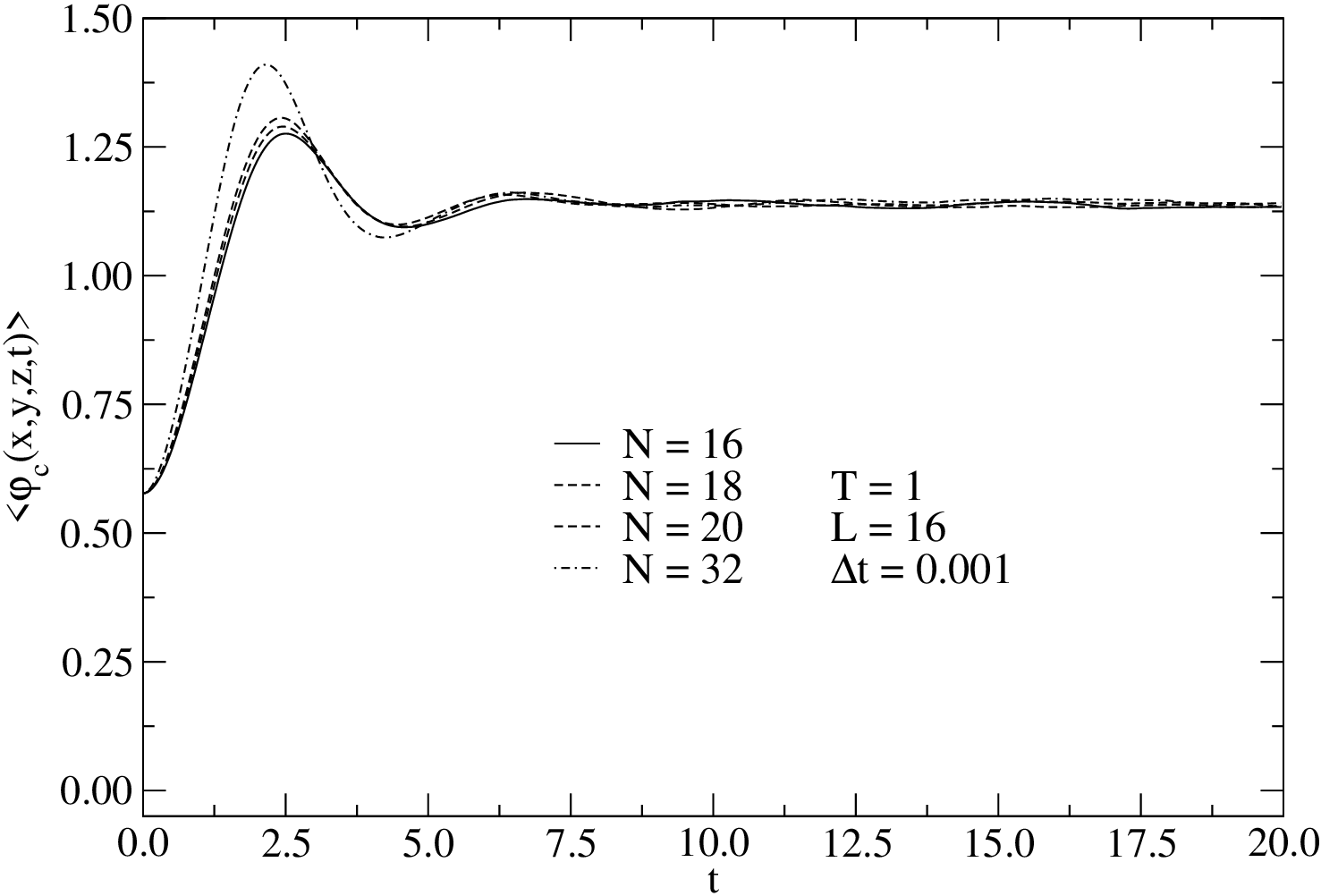}
\end{center}
\caption{Illustration of the lattice size dependence of the solutions
  of the GLL equation for different lattice spacings $L/N$ (left) and the
  effect of counterterms (right). }
\label{fig:latt_indep}
\end{figure}
Equilibrium solutions of the GLL equation insensitive to lattice spacing can
be obtained by the introduction of counterterms in the effective potential.
Since the classical theory in three spatial dimensions is
super-renormalizable, only two terms, corresponding to the first one-loop and
the the last two-loop diagrams terms shown in (\ref{Gamma}), are divergent.
The divergent parts of these graphs can be isolated using a lattice
regularization and then subtracted from the effective potential in the GLL
equation.  Explicit expressions for these counterterms can be found in
\cite{farakosetall}.
Once these are included, one obtains the results shown in {}Fig.~1, where one
can see on the right panel that equilibrium solutions insensitive to lattice
spacing are obtained. Since the counterterms are calculated with the
equilibrium partition function, some sensitivity to lattice spacing remains
for short times.
\begin{figure}[ht]
\begin{center}
  \includegraphics[scale=0.2725]{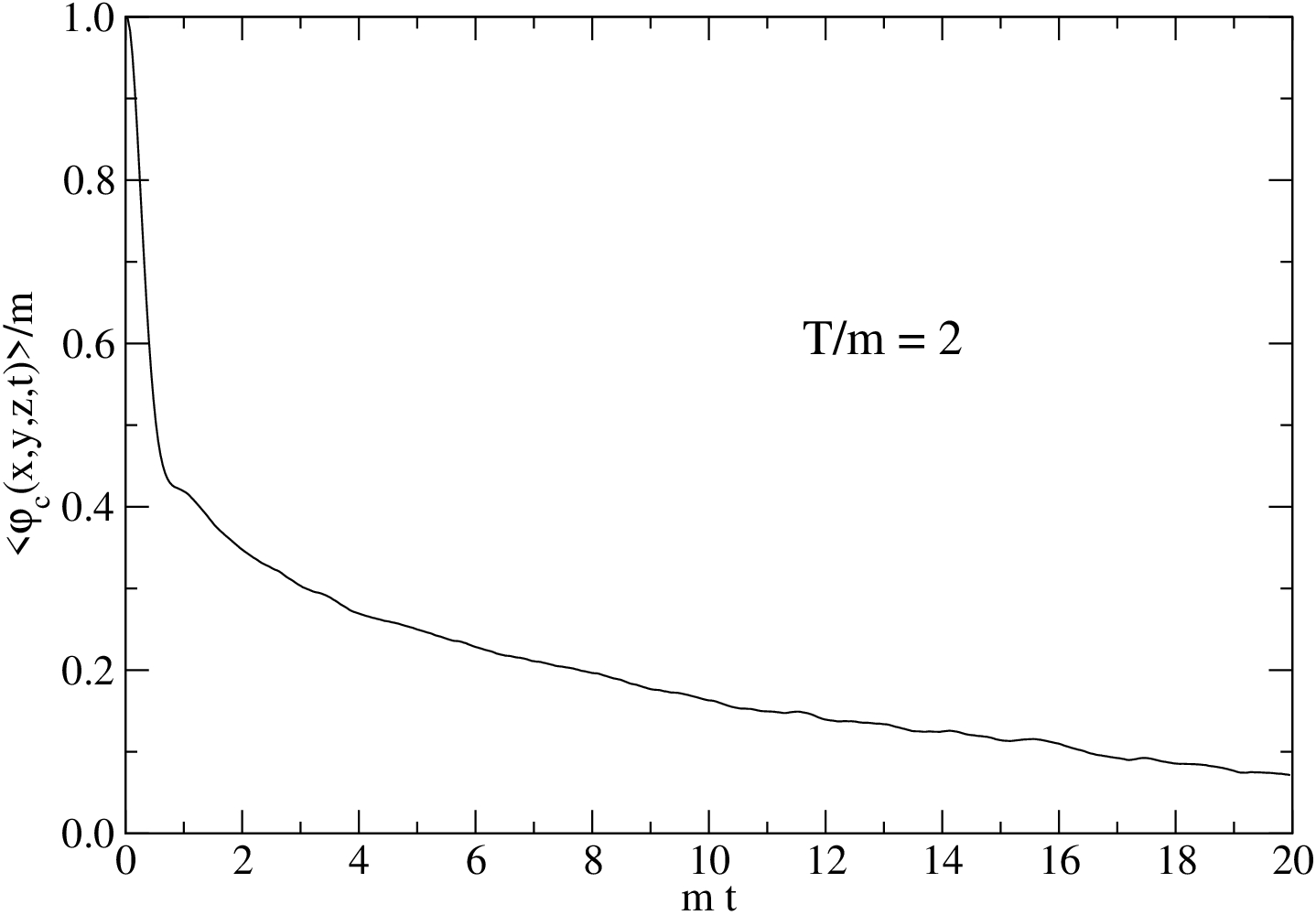}\hspace{0.15cm}
  \includegraphics[scale=0.2725]{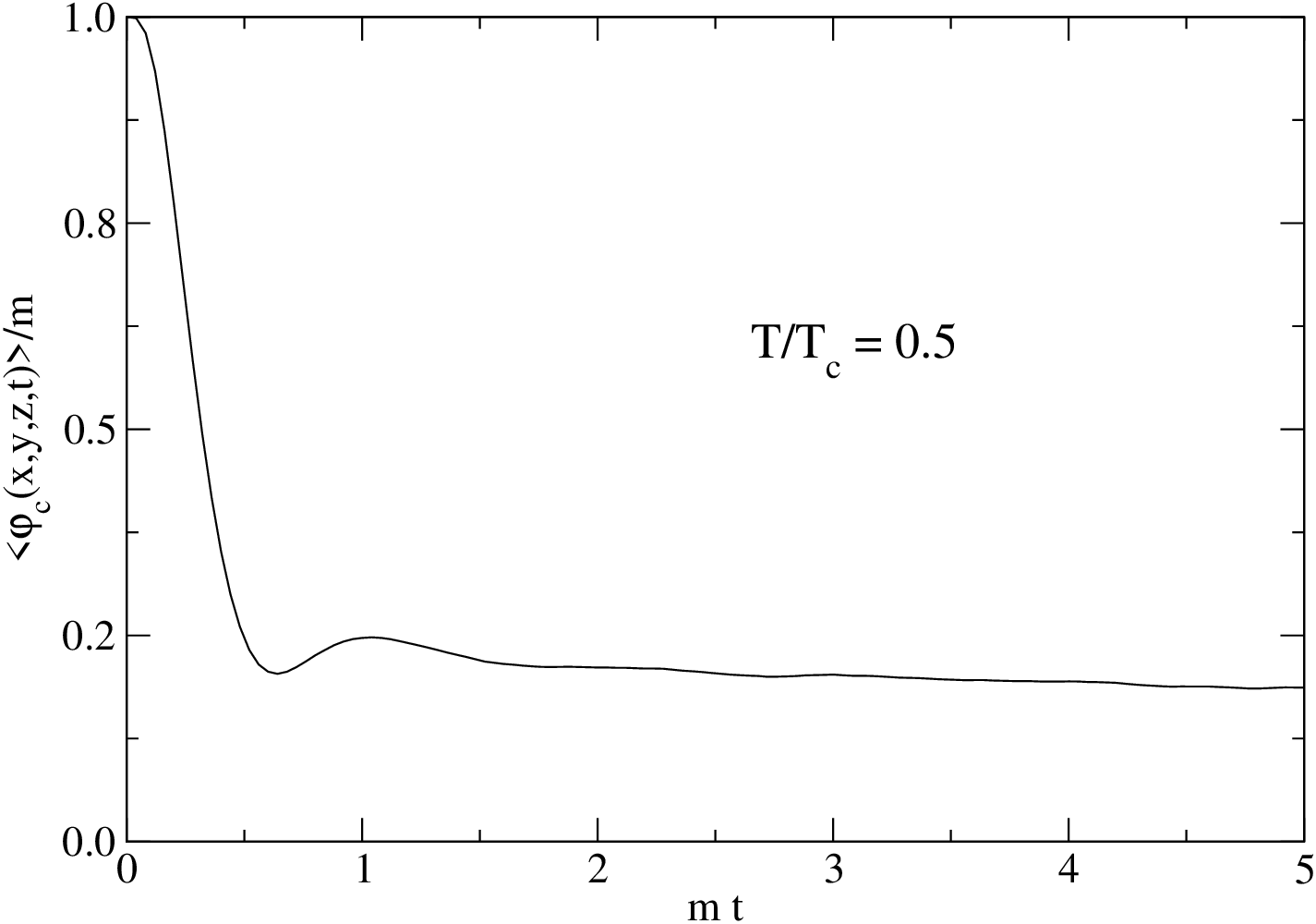}
\end{center}
\caption{Results for the volume average of the order parameter in the
  symmetric (left) and broken (right) phases. }
\end{figure}
Our results for the symmetric and broken phases of the effective GLL equations
for the $\lambda \phi^4$ theory are shown in {}Fig. 2. Since our results are
obtained under the conditions of $\lambda$ small (in Fig. 2 $\lambda = 0.25$)
and for $T/m \gg 1$, the double-well in the broken phase is very shallow and
therefore the thermal fluctuations are large. The field then quickly moves to
a lower vacuum expectation value, but still different from zero, since the
temperature is still smaller than the Ginzburg temperature for the transition.

In conclusion, we have  studied the nonequilibrium dynamics for a
scalar field background configuration. The effective
equation of motion describing the dynamics can be obtained entirely from
microscopic considerations and it is in a form of a Langevin equation with
multiplicative noise and field dependent dissipation terms. It is
suitable to the study of the nonequilibrium dynamics of fields using
standard Langevin dynamics methods on the lattice. Further details  and a
more extended discussion will be presented elsewhere \cite{FSKR}.

Work partially financed by the Brazilian agencies CNPq, FAPERJ and FAPESP.

\end{document}